\documentclass[a4paper,11pt]{article}
\usepackage{pos}

\usepackage{titlesec}
\titlespacing*{\section}      {0pt}{1.2ex plus 0.5ex minus 0.3ex}{0.8ex plus 0.2ex}
\titlespacing*{\subsection}   {0pt}{1.0ex plus 0.4ex minus 0.2ex}{0.5ex plus 0.1ex}
\titlespacing*{\subsubsection}{0pt}{0.8ex plus 0.3ex minus 0.2ex}{0.4ex plus 0.1ex}

\title{Lattice determination of the higher-order hadronic vacuum polarization contribution to the muon $g-2$}
\ShortTitle{Lattice determination of the higher-order HVP contribution to the muon $g-2$}

\author*[a]{Arnau Beltran}
\author[b,c]{Alessandro Conigli}
\author[d]{Simon Kuberski}
\author[a,b,d]{Harvey B. Meyer}
\author[a,e]{Konstantin Ottnad}
\author[a,b,c]{Hartmut Wittig}
\affiliation[a]{\small \textit{PRISMA}$^{++}$ Cluster of Excellence and Institut f\"{u}r Kernphysik, Johannes Gutenberg-Universit\"{a}t Mainz \\ 55099 Mainz, Germany}
\affiliation[b]{\small Helmholtz-Institut Mainz, Johannes Gutenberg-Universit\"{a}t Mainz \\ 55099 Mainz, Germany}
\affiliation[c]{\small GSI Helmholtz Centre for Heavy Ion Research \\ 64291 Darmstadt, Germany}
\affiliation[d]{\small Theoretical Physics Department, CERN \\ 1211 Geneva 23, Switzerland}
\affiliation[e]{\small  Helmholtz-Institut f\"ur Strahlen und Kernphysik (Theory), Rheinische Friedrich-Wilhelms-Universit\"at Bonn \\ Nussallee 14-16, 53115 Bonn, Germany}

\emailAdd{abeltran@uni-mainz.de}

\abstract{%
We present the first lattice QCD calculation of the next-to-leading order (NLO) hadronic vacuum polarization (HVP) contribution to the muon anomalous magnetic moment with sub-percent precision. We employ the time-momentum representation combined with the spatially summed vector correlator computed on CLS ensembles with $N_{\mathrm{f}}=2+1$ flavors of $\mathrm{O}(a)$-improved Wilson fermions, spanning six lattice spacings ($0.039$-$0.097$\,fm) and a range of pion masses including the physical value. After accounting for finite-size corrections and isospin-breaking effects, we obtain in the continuum limit $a_\mu^{\mathrm{hvp,\,nlo}} = (-101.57 \pm 0.26_{\rm stat} \pm 0.54_{\rm syst}) \times 10^{-11}$, corresponding to a total relative error of 0.6\%. Our result lies 1.4$\sigma$ below the estimate of the 2025 White Paper update and is two times more precise. It also shows a tension of $4.6\sigma$ with data-driven evaluations based on hadronic cross section measurements prior to the CMD-3 result.

\vspace*{0.5cm}
\begin{flushright}
	MITP-26-017\\
	CERN-TH-2026-080
\end{flushright}
}

\FullConference{The 42nd International Symposium on Lattice Field Theory (LATTICE2025)\\
2-8 November 2025\\
Tata Institute of Fundamental Research, Mumbai, India\\}

\begin{document}

\maketitle

\section{Introduction}
\label{sec:introduction}

The muon anomalous magnetic moment $a_\mu$ is a high-precision observable that constitutes one of the most sensitive probes of physics beyond the Standard Model (SM). The E989 experiment at Fermilab has recently delivered its final result~\cite{Muong-2:2025xyk}, pushing the experimental precision to 124~ppb. The dominant source of uncertainty on the SM side lies in the leading-order (LO) hadronic vacuum polarization (HVP) contribution $a_\mu^{\mathrm{hvp,\,lo}}$, which enters at order $\alpha^2$. According to the 2025 update of the White Paper by the Muon $g$-$2$ Theory Initiative (WP25~\cite{WP25}), the SM prediction carries an error roughly four times larger than the experimental one, almost entirely from this source.
\begin{figure}[t]
  \centering
  \includegraphics[width=\textwidth]{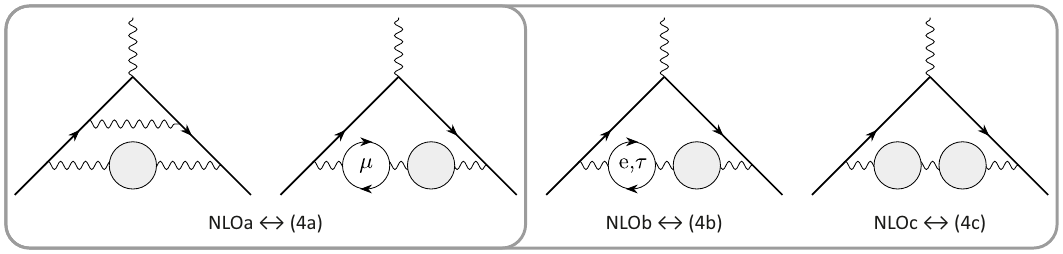}
  \caption{NLO HVP diagrams studied in this work---NLOa (photon lines and muon loops), NLOb (electron and tau loops), and NLOc (two QCD insertions).}
  \label{fig:feynman_diag}
\end{figure}

A critical complication is that the traditional data-driven dispersive approach to $a_\mu^{\mathrm{hvp,\,lo}}$ is sensitive to the specific experimental input used for the dominant $e^+e^-\to\pi^+\pi^-$ cross section. Recent high-statistics measurements by CMD-3~\cite{CMD-3:2023alj, CMD-3:2023rfe} are in tension with earlier results, and the unresolved spread among experiments motivates independent first-principles determinations via lattice QCD. Indeed, several state-of-the-art lattice calculations \cite{Borsanyi:2020mff, Ce:2022kxy, Kuberski:2024bcj, Djukanovic:2024cmq, Boccaletti:2024guq} are now competitive with the data-driven method. As first-principles determinations, they provide a more reliable and independent cross-check of the SM prediction.

For consistency, both the LO and NLO HVP contributions ought to ultimately be evaluated within the same framework. The WP25 estimate for the NLO contribution $a_\mu^{\mathrm{hvp,\,nlo}}$ still relies entirely on the data-driven dispersive approach~\cite{Keshavarzi:2019abf, DiLuzio:2024sps}, which inherits the same experimental tensions affecting the LO determination. A lattice calculation of the NLO contribution is therefore both timely and necessary.

In this contribution we summarize the first lattice QCD determination of $a_\mu^{\mathrm{hvp,\,nlo}}$ with sub-percent precision~\cite{Beltran:2026ofp}. Our calculation extends the framework developed for the LO HVP~\cite{Ce:2022kxy, Kuberski:2024bcj, Djukanovic:2024cmq} to the three classes of NLO diagrams (NLOa, NLOb, NLOc) shown in Fig.~\ref{fig:feynman_diag}. When compared to the LO, the NLOa set of diagram include extra photon-lines and muon-loop corrections and present the dominant negative contribution to the NLO HVP. NLOb considers the other possible lepton-loop corrections, contributing positively and partially canceling NLOa. Lastly, diagram NLOc consists of two QCD insertions and is subleading when compared to the other two.

We employ the space-like representation of the NLO kernel functions \cite{Nesterenko:2021byp,Balzani:2021del} within the time-momentum representation (TMR)~\cite{Bernecker:2011gh,Balzani:2024gmu}, and compute the zero-momentum projected vector correlator on 35 CLS gauge ensembles generated with $N_\mathrm{f}=2+1$ flavors of $\mathrm{O}(a)$-improved Wilson fermions. 

\section{Setup and Formalism}
\label{sec:setup}

\subsection{NLO time-kernels and the NLOa\&b long-distance cancellation}

We relate the NLO HVP contributions to the zero-momentum projection of the electromagnetic correlator $G(t)$ in the TMR through
\begin{equation}
  a_\mu^{\mathrm{hvp},\,(i)} = \left(\frac{\alpha}{\pi}\right)^3\,\int_0^\infty dt_1\dots dt_m\, \tilde{f}^{(i)}(\hat{t}_1,\dots,\hat{t}_m)\,G(t_1)\times\dots\times G(t_m)\, ,
  \label{eq:main_eq}
\end{equation}
for $(i)=(4a)$, $(4b)$ and $(4c)$, where $m=1$ for NLOa and NLOb and $m=2$ for NLOc. The space-like NLO time-kernel functions $\tilde{f}^{(i)}$ entering Eq.~\eqref{eq:main_eq} are not available in closed form but can be obtained from the time-like kernels $\hat{f}^{(i)}$ via a one-dimensional integral transform~\cite{Nesterenko:2021byp, Balzani:2021del, Balzani:2024gmu}. For lattice applications, small-$\hat{t}$ polynomial expansions are sufficient, achieving absolute accuracy better than $10^{-8}$ up to $t\lesssim 7\,\mathrm{fm}$.  For NLOa, the leading expansion coefficient is logarithmically enhanced, $C_4^{(4a)}\propto\gamma_E+\ln\hat{t}$, producing $\mathrm{O}(a^2\ln^2 a)$ lattice artifacts that demand dedicated short-distance treatment (see Sec.~\ref{sec:strategy}). For NLOb, a double expansion in $\hat{t}$ and the mass ratio $M_e=m_e/m_\mu$ is employed. The NLOc kernel admits a full analytic solution and is detailed in the companion paper~\cite{Beltran:2026ofp}.

\begin{figure}[t]
  \centering
  \includegraphics[scale=0.4]{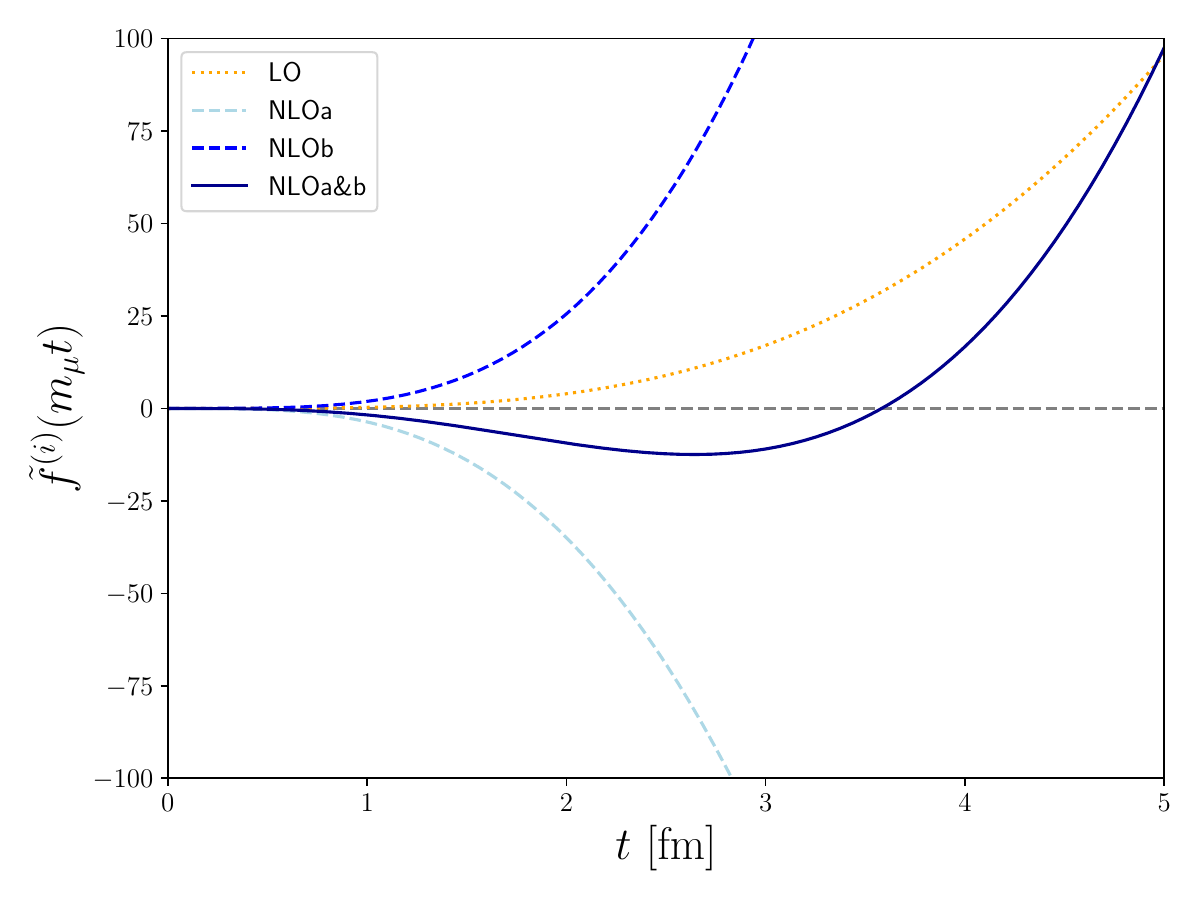}
  \caption{NLOa (dashed), NLOb (dotted), and their sum NLOa\&b (solid blue) time-kernels, compared to the LO kernel (dotted yellow). The cancellation beyond $t\approx1.5\,\mathrm{fm}$ strongly suppresses the long-distance contribution to the total NLO integrand.}
  \label{fig:kernel_comparision}
\end{figure}

A key structural feature of the NLO contribution is the strong cancellation between the NLOa and NLOb time-kernels at large Euclidean time, illustrated in Fig.~\ref{fig:kernel_comparision}: their sum is heavily suppressed relative to the individual contributions and passes through zero at $t\approx3.6\,\mathrm{fm}$.  As a result, the combined NLOa\&b integrand is far less sensitive than the LO integrand to the dominant long-distance systematics---statistical noise, finite-volume effects, isospin-breaking corrections, and scale-setting uncertainties. This cancellation is the single most important feature enabling sub-percent precision and is exploited systematically throughout the analysis.

\subsection{Lattice ensembles, discretization, and extrapolation}

We compute the vector correlator $G(t)$ on 35 gauge ensembles generated by the CLS effort~\cite{Bruno:2014jqa, Bali:2016umi} using a tree-level Symanzik-improved gauge action and $N_\mathrm{f}=2+1$ flavors of non-perturbatively $\mathrm{O}(a)$-improved Wilson fermions~\cite{Bulava:2013cta}. The ensembles cover six lattice spacings $a\approx 0.039$--$0.097\,\mathrm{fm}$ and pion masses from $\sim430\,\mathrm{MeV}$ down to the physical point, following two chiral trajectories: one at approximately fixed $(m_K^2+\tfrac{1}{2}m_\pi^2)$ and a second with near-physical strange quark mass. The scale is set via the gradient-flow observable $\sqrt{t_0}$, using $\sqrt{t_0^{\mathrm{ph}}}=0.1440(7)\,\mathrm{fm}$~\cite{t0madrid}. Full ensemble details are given in the companion paper~\cite{Beltran:2026ofp}.

To assess systematic uncertainties from current discretization, we employ two implementations of the electromagnetic current — local (l) and point-split conserved (c) — each $\mathrm{O}(a)$-improved with coefficient $c_V(g_0)$ determined via two independent non-perturbative methods~\cite{Harris:2025xvk, Heitger:2020zaq}, yielding four independent data sets per ensemble.

Observables are extrapolated to the physical point through a global chiral-continuum fit parametrized in terms of dimensionless combinations of $t_0$, $m_\pi$, and $m_K$. The fit space includes higher-order lattice-spacing and chiral terms, mixed cross terms, and allows for non-zero anomalous-dimension effects. A final result is obtained by model averaging over all ans\"{a}tze, weighted by the Akaike Information Criterion (AIC)~\cite{Akaike:1998zah, Jay:2020jkz}; the weighted spread enters the systematic error budget. The window decomposition of the NLOa and NLOb integrands into short-distance (SD), intermediate-distance (ID), and long-distance (LD) regions follows the standard construction of Ref.~\cite{RBC:2018dos}, with parameters $t_1=0.4\,\mathrm{fm}$, $t_2=1.0\,\mathrm{fm}$, and $\Delta=0.15\,\mathrm{fm}$. Each window is analyzed independently before being summed. Diagram NLOc, being subleading, is treated without any window splitting. To guard against unconscious bias, a multiplicative blinding factor was applied to the long-distance window and lifted only after the analysis was fully frozen.

\section{Analysis Strategy}
\label{sec:strategy}

\subsection{Short-distance window}

The SD window concentrates the most severe lattice artifacts. The logarithmically enhanced $\mathrm{O}(a^2\ln a)$ effects familiar from the LO HVP~\cite{Kuberski:2024bcj} are further compounded in NLOa by $\mathrm{O}(a^2\ln^2\! a)$ terms arising from photon lines, which cannot be reliably constrained by a global continuum extrapolation alone. We eliminate these terms by subtracting the leading kernel behavior, replacing $\Theta_{\mathrm{SD}}(t)\tilde{f}^{(i)}(\hat t)$ with a modified kernel $\tilde{f}^{(i)}_{\mathrm{sub}}(\hat t;Q)$ that removes the dominant $\hat{t}^4$ contribution. The subtracted piece is restored perturbatively as a function of an auxiliary spacelike virtuality $Q$, using the vacuum polarization computed in perturbative QCD to $\mathrm{O}(\alpha_s^4)$~\cite{Chetyrkin_pQCD}. Physical results are independent of $Q$, and we adopt $Q=5\,\mathrm{GeV}$ as our default, where logarithmically enhanced artifacts are negligible and perturbation theory is well controlled. A tree-level multiplicative improvement of the isovector correlator~\cite{Kuberski:2024bcj} further tames residual cutoff effects, and the fit space includes quartic $a^4$ terms for SD observables.

We additionally exploit the approximate $\mathrm{SU}(3)_{\mathrm{f}}$ symmetry restored at high energies. Rather than extrapolating the isoscalar and strange channels independently, we compute only the suppressed differences $\Delta_{ls}(a_\mu) \equiv a_\mu^{8,8} - a_\mu^{3,3}$ and $\Delta_{ls}^{\mathrm{conn}}(a_\mu)$ relative to the isovector channel, reducing the sensitivity to discretization effects and improving the precision of the isoscalar and strange determinations. An analogous strategy is employed for the charm channel.

\subsection{Long-distance window and noise control}

The LD window is dominated by the exponential signal-to-noise problem of the vector correlator. We address this through three complementary methods, following the approach of Ref.~\cite{Djukanovic:2024cmq}: (i) low-mode averaging (LMA)~\cite{Giusti:2004yp, DeGrand:2004qw}, applied to ensembles with $m_\pi \lesssim 280\,\mathrm{MeV}$, which provides a substantially improved estimator for the light-connected correlator by treating the low modes of the Dirac operator exactly; (ii) a spectral reconstruction of the isovector correlator tail for two close-to-physical-point ensembles (E250 and D200), using energy levels extracted from a GEVP analysis~\cite{Djukanovic:2024cmq}, which sharply reduces the uncertainty in the chiral extrapolation of the LD window; and (iii) the standard bounding method~\cite{RBC:2018dos}, which constrains the correlator at large $t$ using only the ground-state energy and the value of $G(t)$ at a cutoff time $t_c$.

Crucially, by always working with the combined NLOa\&b integrand $\tilde{f}^{(4a)}+\tilde{f}^{(4b)}$, whose kernel is strongly suppressed relative to the individual diagrams beyond $t\sim 1.5\,\mathrm{fm}$ relative to the individual diagrams (see Fig.~\ref{fig:kernel_comparision}), the practical weight of the LD window in the total uncertainty budget is greatly diminished compared to what either NLOa or NLOb alone would require.

\subsection{Finite-volume corrections}

Finite-volume (FV) effects are most significant in the LD window and largest in the isovector channel, where they are driven by two-pion intermediate states. We correct for them using a two-step procedure adapted from Ref.~\cite{Djukanovic:2024cmq}. First, each ensemble is corrected to a common reference volume ($m_\pi L)^\mathrm{ref}\approx4.29$ using the Hansen--Patella (HP) formalism~\cite{HP1,HP2} at small $t$, transitioning to the Meyer--Lellouch--L\"{u}scher (MLL) method~\cite{MLL} at larger $t$, where the two-particle quantization condition is more reliably implemented. Second, the residual correction from the reference volume to infinite volume is evaluated in the continuum limit using NNLO chiral perturbation theory~\cite{hvp_milc_ChPT}, with a 10\% uncertainty assigned to this step.

\subsection{NLOc diagram}

The NLOc contribution, involving two separate QCD insertions, enters the TMR as a double time integral over $\tilde{f}^{(4c)}(\hat{t},\hat{\tau})\,G(t)\,G(\tau)$. Its magnitude is roughly 25 times smaller than NLOa\&b, so no window decomposition is needed. FV corrections are applied directly to infinite volume using the HP\&MLL method---the intermediate $L_\mathrm{ref}$ step is unnecessary at this level of precision. The bilinear structure of the integrand introduces crossed FV terms, i.e. corrections to $G^a(t)\,G^b(\tau)$ from simultaneous FV shifts in both correlators, which we retain for consistency.

\subsection{Isospin-breaking corrections}

All CLS ensembles are generated in the isospin-symmetric QCD limit. Conversion to the physical theory requires electromagnetic (EM) and strong isospin-breaking (IB) corrections. Because generating the required data for a lattice evaluation of these effects is beyond the scope of this project, we resort instead to phenomenological estimates via the spacelike representation of the HVP scalar function. EM corrections are modelled using vector meson dominance (VMD)~\cite{Volodymyr1, Volodymyr2}, dominated by the charged--neutral pion mass splitting. Strong IB corrections are estimated from the $(3,8)$ component of the vacuum polarization~\cite{Volodymyr3Erb_SIB}. A conservative 50\% uncertainty is assigned to each contribution. The same cancellation that suppresses the long-distance noise for NLOa\&b also operates here: the EM and strong IB corrections largely cancel between NLOa and NLOb at small spacelike momenta, yielding a fortuitously small and well-controlled total IB shift $\Delta^{\mathrm{IB}}a_\mu^{\mathrm{hvp,\,nlo}} = 0.06(27)\times10^{-11}$ for the combined NLO contribution.

\section{Results}
\label{sec:results}

\subsection{isoQCD results for NLOa, NLOb, and NLOc}

The first three entries in table~\ref{tab:isoQCD_result} collect the isoQCD results for NLOa, NLOb and their combination broken down by time window. The extrapolations for all three are performed independently; the excellent agreement between the directly extrapolated NLOa\&b and the sum of the separately extrapolated NLOa and NLOb provides a non-trivial internal consistency check. We also show the total estimate for NLOc.
\begin{table}[t]
  \centering
  \small
  \renewcommand{\arraystretch}{1.1}
  \setlength{\tabcolsep}{6pt}
  \begin{tabular}{c | r@{.}l r@{.}l r@{.}l | r@{.}l}
    \noalign{\smallskip}\hline\noalign{\smallskip}
    & \multicolumn{2}{c}{SD}
    & \multicolumn{2}{c}{ID}
    & \multicolumn{2}{c|}{LD}
    & \multicolumn{2}{c}{Total} \\
    \noalign{\smallskip}\hline\noalign{\smallskip}
    NLOa    & $-$34&78(19) & $-$82&83(30) &  $-$99&68(1.11) & $-$217&28(1.23) \\
    NLOb    &  10&72(5)   &  36&88(14)   &   64&42(83)  &  112&02(88) \\
    \noalign{\smallskip}\hline\noalign{\smallskip}
    NLOa\&b & $-$24&05(13) & $-$45&95(16) &  $-$35&28(31)  & $-$105&29(40) \\
    \noalign{\smallskip}\hline\noalign{\smallskip}
    NLOc & \multicolumn{2}{c}{-} & \multicolumn{2}{c}{-} & \multicolumn{2}{c|}{-} & 3&78(9) \\
    \noalign{\smallskip}\hline\noalign{\smallskip}
  \end{tabular}
  \caption{isoQCD results for NLOa, NLOb, and their combination by window, as well as the total estimate for NLOc.
    Stat.\ and syst.\ uncertainties are combined in quadrature.
    All values in units of $10^{-11}$.}
  \label{tab:isoQCD_result}
\end{table}

The window breakdown reveals the central role of the NLOa\&b cancellation: while the LD windows of NLOa and NLOb individually carry uncertainties of $1.11\times10^{-11}$ and $0.83\times10^{-11}$ respectively, the combined LD uncertainty shrinks to $0.31\times10^{-11}$, and the LD window accounts for only one third of the total NLOa\&b estimate---a sharp contrast with the LO HVP.

For NLOc, the dominant contribution is from the isovector double insertion $a_\mu^{3,3-3,3}$, while the largest uncertainty comes from the isovector--isoscalar crossed term $(2/3)\,a_\mu^{3,3-8,8}$. After combining all flavor channels we obtain a 2.4\% precision for the isoQCD estimate of this diagram.

\subsection{Subleading contributions}

Several corrections beyond the main calculation are required for the physical result. The bottom-quark contribution to NLOa and NLOb is estimated perturbatively via a time-like dispersion integral exploiting the smallness of $\alpha_s(m_b^2)/\pi\approx0.07$, yielding $(1/9)\,a_\mu^{b,b,\,\mathrm{nlo(a)}} \approx -0.23\times10^{-11}$ and $(1/9)\,a_\mu^{b,b,\,\mathrm{nlo(b)}} \approx +0.048\times10^{-11}$. Charm disconnected contributions are computed on the lattice and are numerically irrelevant. Charm sea-quark quenching effects are estimated phenomenologically and included in the error budget. The tau-loop contribution to NLOb, previously neglected, is estimated using ensemble E250 to be $a_\mu^{\mathrm{hvp,\,nlo(b;\tau)}} = 0.06(3)\times10^{-11}$.

\subsection{Full NLO HVP result and comparison}

Combining the isoQCD NLOa\&b and NLOc results with the isospin-breaking correction and all subleading contributions, we obtain the per-diagram physical results
\begin{equation}
  \begin{aligned}
      a_\mu^{\mathrm{hvp,\,nlo(a)}} & = -216.83(81)(87)(99)(81)(34)(31)[1.81] \times 10^{-11} \, , \\
      a_\mu^{\mathrm{hvp,\,nlo(b)}} & = \phantom{-}111.51(55)(64)(64)(58)(16)(25)[1.24] \times 10^{-11} \, , \\
      a_\mu^{\mathrm{hvp,\,nlo(a\&b)}} & = -105.33(28)(28)(35)(24)(18)(5)[61] \times 10^{-11} \, , \\
      a_\mu^{\mathrm{hvp,\,nlo(c)}} & = \phantom{-00}3.75(6)(6)(3)(3)[10] \times 10^{-11} \, , \\
  \end{aligned}
  \label{eq:final_result_diag}
\end{equation}
where the individual errors are, in order: statistical, model-average systematic, scale setting ($t_0$), isospin breaking, FV correction in the continuum, and charm sea-quark effects. The total, combining NLOa\&b and NLOc, is our main result:
\begin{equation}
  \phantom{-}a_\mu^{\mathrm{hvp,\,nlo}} = -101.57(26)(29)(31)(27)(18)(5)[59] \times 10^{-11} \, .
  \label{eq:final_result}
\end{equation}
The total uncertainty of $0.59\times10^{-11}$ corresponds to a relative precision of \textbf{0.6\%}, matching the precision in some data-driven evaluations and improving by a factor two the precision quoted in the WP25 estimate. The uncertainty is dominated in roughly equal parts by the LD statistical noise, the model-average spread, the scale-setting uncertainty from $t_0$, and the isospin-breaking correction.

\begin{figure}[t]
  \centering
  \includegraphics[width=\textwidth]{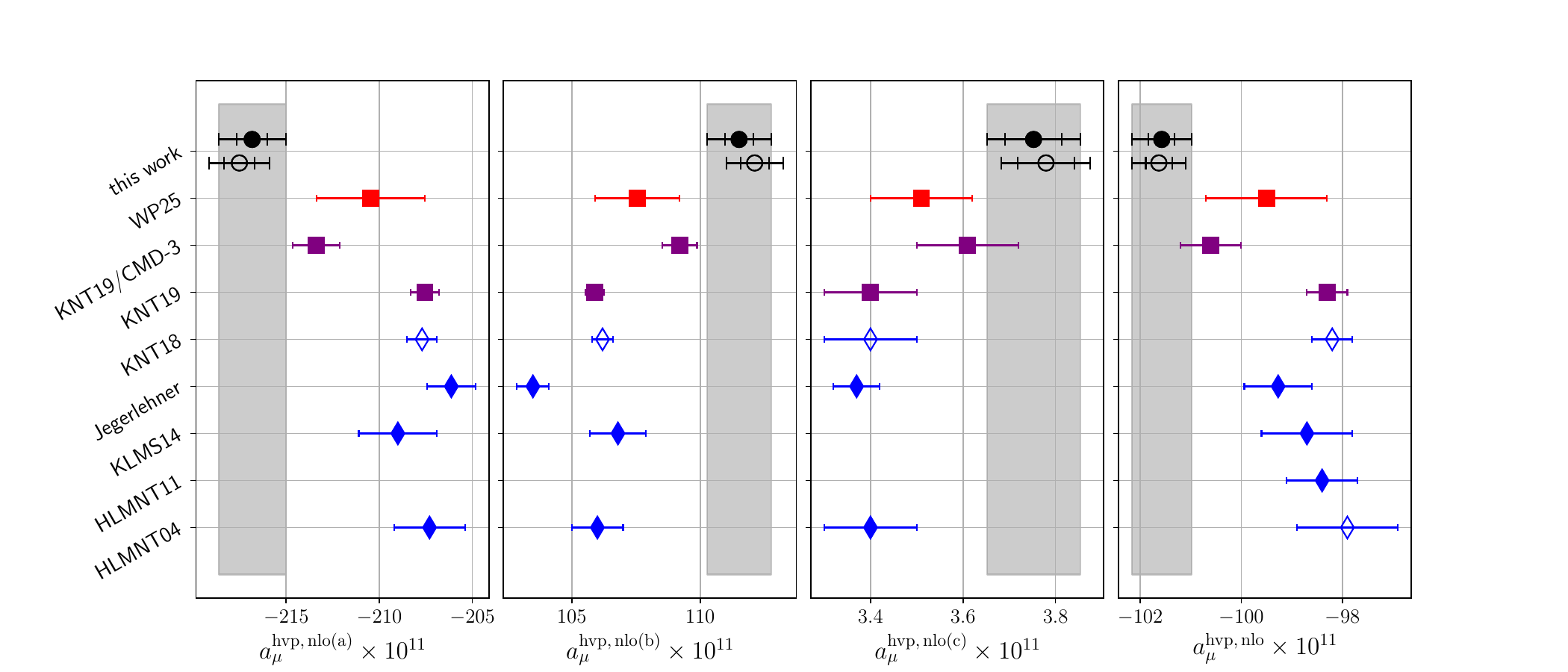}
  \caption{Comparison of our results (filled black point: physical; open black point: isoQCD) with data-driven evaluations. Red squares show the WP25 average~\cite{WP25} together with its KNT19~\cite{Keshavarzi:2019abf} and KNT19/CMD-3~\cite{DiLuzio:2024sps} inputs (purple). Older data-driven results are shown in blue rhombuses~\cite{Hagiwara:2006jt,HLMNT11,KLMS14,Jegerlehner:2017gek,KNT18}.}
  \label{fig:results}
\end{figure}

Figure~\ref{fig:results} compares Eqs.~\eqref{eq:final_result_diag} and~\eqref{eq:final_result} with a range of data-driven evaluations. As shown in the right-most panel, our result lies $1.4\sigma$ below the WP25 average of $(-99.6\pm1.3)\times10^{-11}$~\cite{WP25} and shows a $4.6\sigma$ tension with the KNT19 determination~\cite{Keshavarzi:2019abf}, which does not include the CMD-3 measurement. This pattern mirrors the analogous tension observed in $a_\mu^{\mathrm{hvp,\,lo}}$ between lattice and pre-CMD-3 data-driven evaluations.

\section{Conclusion}
\label{sec:conclusions}

We have presented the first lattice QCD calculation of the NLO hadronic vacuum polarization contribution to the muon anomalous magnetic moment with sub-percent precision. Using 35 CLS ensembles spanning six lattice spacings and a range of pion masses down to the physical point, and working entirely within the time-momentum representation, we obtain
\begin{equation*}
  \boxed{
  a_\mu^{\mathrm{hvp,\,nlo}}
  = (-101.57 \pm 0.26_{\mathrm{stat}} \pm 0.54_{\mathrm{syst}})
    \times 10^{-11}\,,}
\end{equation*}
with a total relative error of 0.6\%, a factor of two more precise than the WP25 estimate. This result is consistent with but below the WP25 value at the $1.4\sigma$ level, and shows a $4.6\sigma$ tension with the KNT19 data-driven evaluation that does not incorporate the CMD-3 measurement---fully in line with the pattern of deviations seen for the LO HVP between lattice and pre-CMD-3 dispersive results.

The key feature that allows us to achieve a significantly better precision than for the LO HVP is the strong cancellation between the NLOa and NLOb time-kernels beyond $t\approx1.5\,\mathrm{fm}$. By computing the sum NLOa\&b as a single quantity, the long-distance window---which drives statistical noise, finite-volume corrections, and scale-setting sensitivity---contributes only one third of the total central value, compared to roughly two thirds for the LO HVP. This makes the NLO contribution intrinsically easier to determine precisely than the LO, once the formalism is in place.

Our calculation provides the first fully consistent lattice determination of both the LO and NLO HVP contributions, eliminating the methodological inconsistency in the WP25 prediction where only the LO part was taken from lattice QCD. For practical updates, Appendix~D of the companion paper~\cite{Beltran:2026ofp} tabulates the derivatives of all observables with respect to $\sqrt{t_0}$, $m_\pi$, $m_K$, and $m_{D_s}$, allowing the result to be straightforwardly updated as scale-setting inputs improve, without repeating the full analysis.

Going forward, the focus returns to the LO HVP, which continues to dominate the SM uncertainty budget.

\hfill\break
\noindent \textbf{Acknowledgments:} 
We are grateful to our colleagues in the CLS initiative for sharing ensembles.
Calculations were performed on the HPC clusters at the Helmholtz Institute
Mainz, Johannes Gutenberg-Universit\"{a}t Mainz, J\"{u}lich Supercomputing
Centre (JSC), H\"{o}chstleistungsrechenzentrum Stuttgart (HLRS), and Leibniz
Supercomputing Centre (LRZ). We gratefully acknowledge the support of the
Gauss Centre for Supercomputing (GCS) and the John von Neumann-Institut
f\"{u}r Computing (NIC) via projects HMZ21, HMZ23 and HINTSPEC at JSC,
GCS-HQCD and GCS-MCF300 at HLRS and LRZ, and NHR-SW of JGU Mainz
(project NHR-Gitter). This work was supported by the Deutsche
Forschungsgemeinschaft (DFG) through the Collaborative Research Center
1660, research unit FOR~5327 (Project No.~458854507), grant
HI~2048/1-2 (Project No.~399400745), and the Cluster of Excellence
PRISMA+ (EXC~2118/1, Project No.~390831469). This project has
received funding from the European Union's Horizon Europe programme
under the Marie Sk\l{}odowska-Curie grant agreement No.~101106243.


\newpage

\bibliographystyle{JHEP}
\bibliography{ref}

\end{document}